\documentstyle{black}
\input{psfig}

\def\Mpc{\, h^{-1} \, {\rm Mpc}}

\begin{document}

\title{%
HOW SMOOTH IS THE UNIVERSE ON LARGE SCALES ?}

\author{Ofer LAHAV \\
{\it Institute of Astronomy, Madingley Road, Cambridge CB3 0HA, UK,
lahav@ast.cam.ac.uk}\\ }

\maketitle

\section*{Abstract}

  We review cosmological inference from optical and radio 
  galaxy surveys, the X-Ray 
  Background and the  Cosmic Microwave Background.
  We focus on three topics: 
  (i) First results from the 2dF galaxy redshift survey;
  (ii) Estimation of  cosmological parameters by joint analysis of 
  redshift surveys and CMB data;
  and
  (iii) The validity of the Cosmological Principle, 
   and constraints on the fractal dimension on large scales.
\footnote{Invited talk, to appear in the Proceedings of 
the 49th Yamada conference 
on {\it  Black Holes and High Energy Astrophysics}, 
held in Kyoto in  April 1998, in honour of Prof. H. Sato}

\section{Introduction}

In summarizing a conference in memory of  Yukawa nearly 10 years ago, 
Prof. H. Sato referred to the dark matter problem as  made 
of three `islands' : (i) Astronomical facts concerning galaxies
and the large scale structure of the universe, (ii) the Big Bang model 
and (iii) Particle Physics.
Here we shall focus on the first `island' and discuss briefly 
implications to dark matter models and the validity of the 
Cosmological Principle. 

It is believed by most cosmologists that 
on the very large scales the universe 
is isotropic and homogeneous. 
However, on scales much smaller than the horizon the distribution 
of luminous matter is clumpy.
Surveys such
as CfA, SSRS, IRAS, APM and Las Campanas have yielded useful information on
local structure and on the density parameter  $\Omega$ from
redshift distortion and from comparison with the peculiar velocity
field.  Together with measurements of the Cosmic Microwave Background
(CMB) radiation and gravitational lensing the redshift surveys provide
major probes of the world's geometry and the dark matter.

In spite of the rapid progress
two gaps remain in our understanding of 
the density fluctuations as a function of scale:
(i) It is still unclear how to relate the distributions of 
    galaxies and mass;  
(ii) Little is known about fluctuations 
on  intermediate scales 
between these of local galaxy surveys ($\sim 100 h^{-1} $ Mpc)
and the scales probed by COBE ($\sim 1000 h^{-1} $ Mpc). 

Another related but unresolved issue is the value of the density parameter
$\Omega$.  Putting together different cosmological observations, the
derived values seem to be inconsistent with each other.  
Taking into account moderate
biasing, the redshift and peculiar velocity data on large scales yield
$\Omega \approx 0.3 -1.5$, with a trend towards the popular 
value $\sim 1$
(e.g. Dekel 1994; Strauss \& Willick 1995 for summary of results).
On the other hand, the high fraction of baryons in clusters, combined
with the baryon density from Big Bang Nucleosynthesis suggests $\Omega
\approx 0.2$ (White et al. 1993).  Moreover, an $\Omega=1$ universe is
also in conflict with a high value of the
Hubble constant ($H_0 \approx 70-80$ km/sec/Mpc),
as in this model the universe turns out to be younger
than globular clusters.  A   way out of these problems 
was suggested by 
adding a positive cosmological constant, such that $\Omega + \lambda
=1$, to satisfy inflation.  Two recent observations 
constrain 
$\lambda $ : the observed frequency of lensed quasars is too small,
yielding an upper limit $\lambda <0.65$ 
(e.g. Kochanek 1996),
and the
magnitude-redshift relation for Supernovae type Ia 
(e.g. Perlmutter et al. 1998).
The next decade will
see several CMB experiments (e.g. Planck, MAP, VSA) which promise to
determine (in a model-dependent way) the cosmological parameters to
within  a few percent.
We shall focus here on several issues related to clustering and
cosmological parameters from new surveys.


\section {The 2dF Galaxy Redshift Survey}

Existing optical and IRAS redshift surveys contain $\sim 10^4$ 
galaxies.  
Multifibre technology now allows
us to produce redshift surveys of millions of
galaxies.  Two major surveys have just started.
The American-Japanese 
Sloan Digital Sky Survey (SDSS) will yield images in 5
colours for 50 million galaxies, 
and redshifts for about 1 million galaxies over a quarter of the
sky (Gunn and Weinberg 1995). It will
be carried out using a dedicated 2.5m telescope in New Mexico.  The
median redshift of the survey is ${\bar z} \sim 0.1$.

A complementary Anglo-Australian survey, the 2 degree Field (2dF)
will produce redshifts for 250,000 galaxies brighter than $b_J =19.5^m$
(with median redshift of ${\bar z} \sim 0.1$), selected from the APM catalogue.  The
survey will utilize a new 400-fibre system on the 4m AAT, covering
$\sim 1,700$ sq deg of the sky.  
About 10,000 redshifts have been measured so far (as of June 1998).
A deeper
extension down to $R=21$ for 10,000 galaxies is also planned for the 2dF 
survey. 

The main goals of the 2dF galaxy survey are: 

\hangindent \parindent \hangafter 1
$\bullet$ Accurate measurements of the power spectrum of galaxy
clustering on scales $ > 30 h^{-1}$ Mpc, allowing a direct
comparison with CMB  anisotropy measurements
such as the recently approved  NASA MAP and ESA Planck Surveyor
satellites. The power-spectrum derived from the projected APM galaxies
(see Figure 2) gives an idea about the scales probed by the 2dF
redshift survey. 

\hangindent \parindent \hangafter 1
$\bullet$ Measurement of the distortion of the clustering pattern in
redshift space providing constraints on the cosmological density
parameter $\Omega$ and the spatial distribution of dark matter.

\hangindent \parindent \hangafter 1
$\bullet$ Determination of variations in the spatial and velocity
distributions of galaxies as a function of luminosity, spectral type and
star-formation history, providing important constraints on
models of galaxy formation.

\hangindent \parindent \hangafter 1
$\bullet$ Investigations of the morphology of galaxy clustering and
the statistical properties of the fluctuations, {\it e.g.} whether the
initial fluctuations are Gaussian as predicted by inflationary models
of the early universe.

\hangindent \parindent \hangafter 1
$\bullet$ A study of clusters and groups of galaxies in the redshift
survey, in particular the measurement of infall in clusters and
dynamical estimates of cluster masses at large radii.

\hangindent \parindent \hangafter 1
$\bullet$ Application of novel techniques 
(e.g. Principal Component Analysis 
and Artificial Neural Networks) 
to classify the uniform
sample of $250,000$ spectra, thereby obtaining
a comprehensive inventory of galaxy types as a function of spatial
position within the survey.

\begin{figure}
\centerline{\psfig{figure=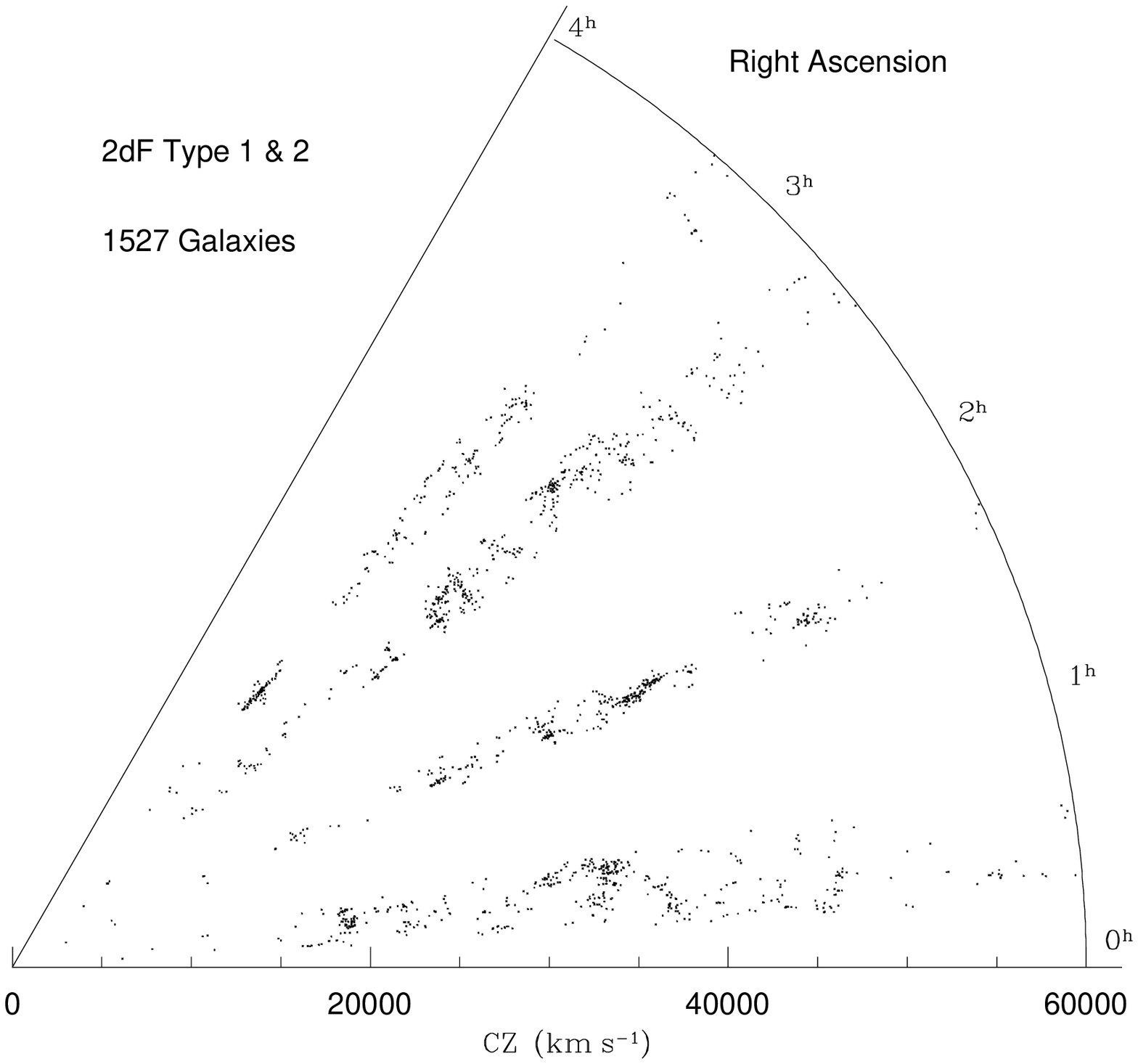,height=7cm,width=8cm}}
\centerline{\psfig{figure=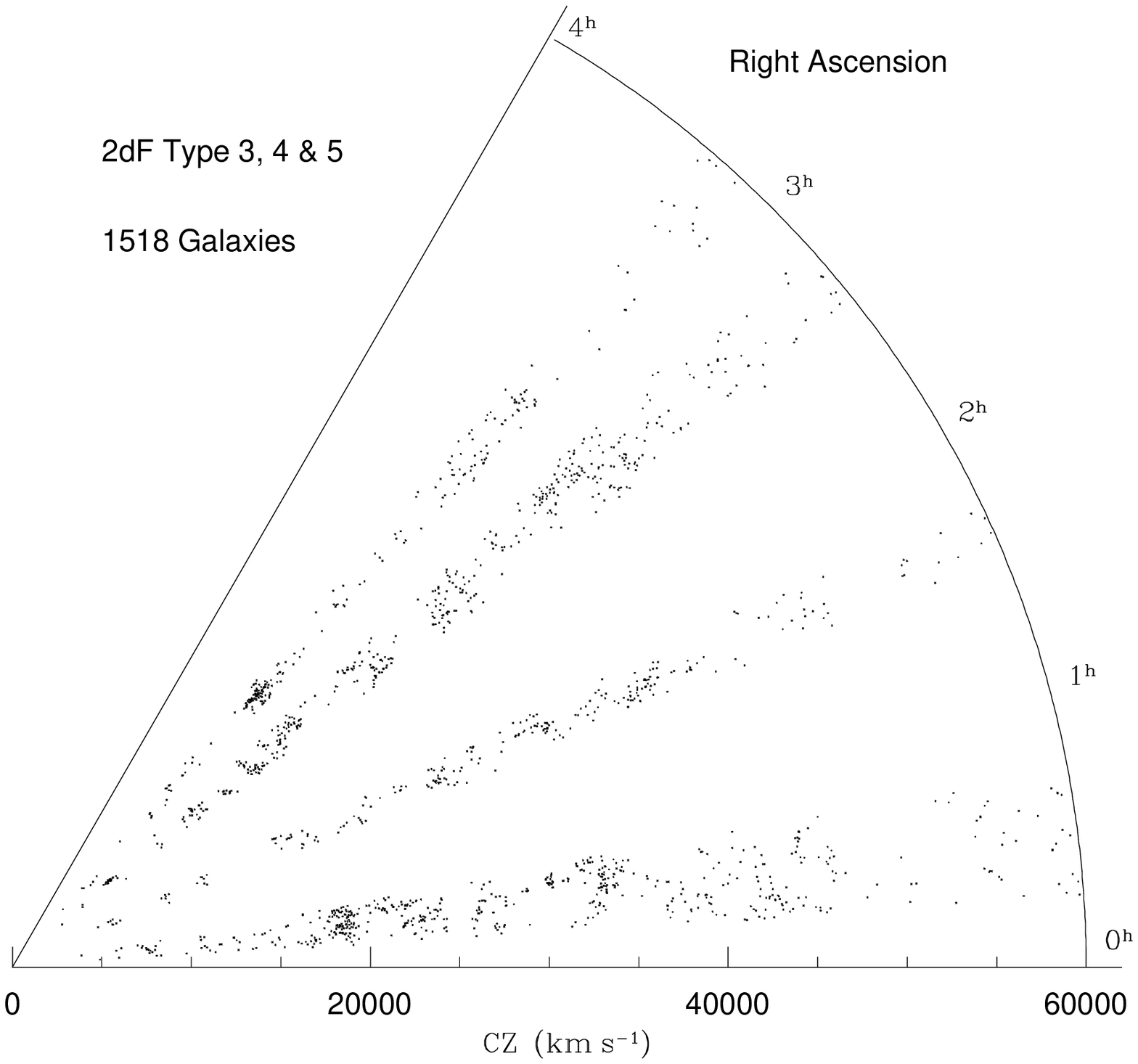,height=7cm,width=8cm}}
\caption{Cone plots of the 2dF galaxies 
 with measured redshift, 
split into `red'  (spectral Types 1 \& 2) and 'blue' 
(spectral Types 3, 4 \& 5) samples. 
The `red' galaxies are more strongly clustered (from Folkes et al. 1998).} 
\label{}
\end{figure}

Figure 1 (from Folkes et al. 1998) shows cone plots of 
redshift-space  distribution for a subset of $\sim 3000$ 2dF galaxies.
The galaxies were classified 
according to their spectra by Principal Component Analysis 
(Folkes, Lahav \& Maddox 1996) 
and then divided into two groups of nearly equal numbers.
The `red' (early type)  galaxies 
do appear more clustered, with evidence for 'finger-of-God'
effects caused by the velocity dispersion of galaxy clusters, 
while the `blue' (late-type) galaxies show a more uniform distribution, 
although clustering is still evident.
This is in qualitative agreement with the well-known
morphology-density relation.
Quantifying these differences and comparing them with the predictions
of models will be a major focus of the future analysis of
the 2dF galaxy survey.
For more details on the 2dF galaxy survey see

{\tt http://msowww.anu.edu.au/$\sim$colless/2dF/}

 \section {Probes at High Redshift}

 The big new surveys  (SDSS, 2dF) 
 will only probe a median redshift ${\bar z} \sim 0.1$.
 It  remains crucial to probe the density fluctuations at higher $z$,
 and  to fill in the gap between
 scales probed by previous local galaxy surveys and the scales 
 probed by COBE and other CMB experiments.
 Here we discuss the X-ray Background (XRB) and radio sources 
 as probes of the density fluctuations at median redshift $ {\bar z} \sim 1$.
 Other possible high-redshift traces are quasars and clusters of galaxies.

 \subsection{Radio Sources}

Radio sources in surveys have typical median redshift
${\bar z} \sim 1$, and hence are useful probes of clustering at high
redshift. 
Unfortunately, it is difficult to obtain distance information from
these surveys: the radio luminosity function is very broad, and it is
difficult to measure optical redshifts of distant radio sources.
Earlier studies
claimed that  the distribution of radio sources supports the 
'Cosmological Principle'.
However, 
the wide range in intrinsic luminosities of radio sources
would dilute any clustering when projected on the sky.  
Recent analyses  of
new deep radio surveys (e.g. FIRST)
suggest that radio sources are actually  clustered at least as strongly
as local optical
galaxies 
(e.g. Cress et al. 1996; Magliocchetti et al. 1998).
Nevertheless, on the very large scales the distribution of radio sources
seems nearly isotropic. 
Comparison of the measured quadrupole in a radio sample 
in the Green Bank and Parkes-MIT-NRAO
4.85 GHz surveys 
to the theoretically predicted ones (Baleisis et al. 1998)
offers a crude estimate of the fluctuations on scales $ \lambda \sim 600
h^{-1}$ Mpc.  The derived amplitudes are shown in Figure 2 for the two 
assumed Cold Dark Matter (CDM) models.
Given the problems of catalogue matching and shot-noise, these points should be
interpreted at best as `upper limits', not as detections.

\begin{figure}
\protect\centerline{
\psfig{figure=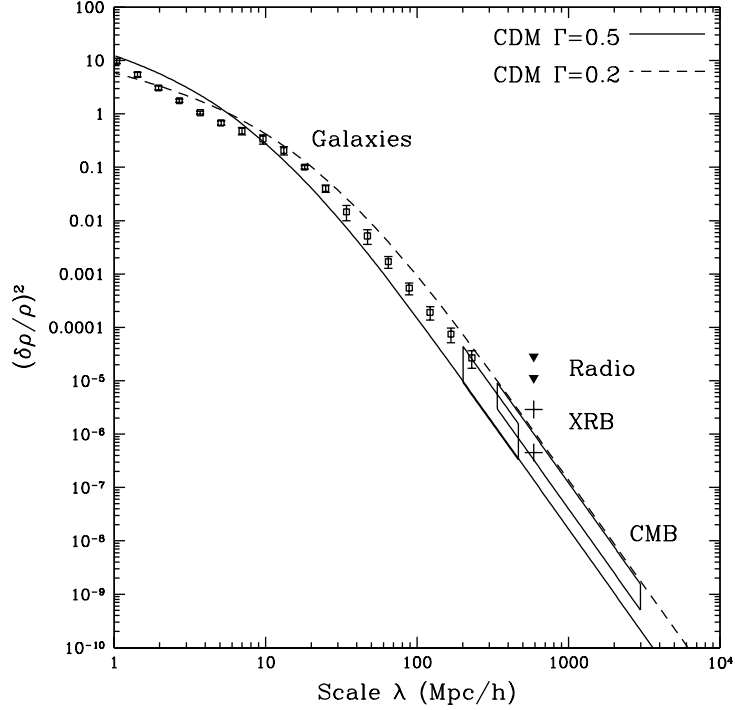,height=4truein,width=4truein}}
\caption[]{
  A compilation of density fluctuations on different scales from
  various observations: a galaxy survey, deep radio surveys, the X-ray
  Background and Cosmic Microwave Background experiments. The
  measurements are compared with two popular Cold Dark Matter models
  (with normalization $\sigma_8=1$ and 
  shape parameters $\Gamma=0.2$ and $0.5$). 
  The Figure shows mean-square density fluctuations $({ {\delta \rho}
    \over \rho })^2 \propto k^3 P(k)$, where $k=1/\lambda$ is the
  wavenumber and $P(k)$ is the power-spectrum of fluctuations.  
  The open squares at
  small scales are estimates from 
  the APM galaxy catalogue (Baugh \& Efstathiou 1994).
  The elongated 'boxes' at large scales represent the COBE
  4-yr (on the right) and
  Tenerife (on the left) CMB measurements  (Gawiser \& Silk
  1998).  The solid triangles and crosses represent amplitudes derived
  from the
  quadrupole of radio sources 
  (Baleisis et al. 1998) and the quadrupole of the XRB 
  (Lahav et al. 1997; Treyer et al. 1998).  
  Each pair of estimates corresponds to assumed shape of the two CDM models.
  (A compilation from Wu, Lahav \& Rees 1998).  }
\end{figure}

 \subsection {The XRB}

 Although discovered in 1962, the origin of
 the X-ray Background (XRB) is still unknown,  
 but is likely
 to be due to sources at high redshift 
 (for review see Boldt 1987; Fabian \& Barcons 1992).
 Here we shall not attempt to speculate on the nature of the XRB sources.
 Instead, we {\it utilise} the XRB as a probe of the density fluctuations at
 high redshift.  The XRB sources are probably
 located at redshift $z < 5$, making them convenient tracers of the mass
 distribution on scales intermediate between those in the CMB as probed
 by COBE, and those probed by optical and IRAS redshift
 surveys (see Figure 2).

The interpretation of the results depends somewhat on the nature of
the X-ray sources and their evolution.  The rms dipole and higher
moments of spherical harmonics can be predicted (Lahav et al. 
1997) in the
framework of growth of structure by gravitational instability from
initial density fluctuations.
By comparing
the predicted multipoles to those observed by HEAO1 
(Treyer et al. 1998)
we estimate the amplitude of fluctuations for an
assumed shape of the density fluctuations 
(e.g. CDM models).  
Figure 2 shows the amplitude of fluctuations derived at the 
effective scale $\lambda \sim 600 h^{-1}$ Mpc probed by the XRB. 
The observed fluctuations in the XRB
are roughly as expected from interpolating between the
local galaxy surveys and the COBE CMB experiment.
The rms fluctuations 
${ {\delta \rho} \over {\rho} }$
on a scale of $\sim 600 h^{-1}$Mpc 
are less than 0.2 \%.

  \section {Is the FRW Metric Valid on Large Scales ?}

The Cosmological Principle was first adopted when observational
cosmology was in its infancy; it was then little more than a
conjecture. Observations could not then probe to significant redshifts, the
`dark matter' problem was not well-established and the Cosmic Microwave
CMB and the XRB were still unknown.  
If the  Cosmological Principle turned out to be invalid 
then the consequences to our understanding of cosmology would be dramatic, 
for example the conventional way of interpreting the age of the universe, 
its geometry and matter content would have to be revised. 
Therefore it is 
important to revisit this underlying assumption in the light of new
galaxy surveys and measurements of the background radiations.
The question of whether the universe 
is isotropic and homogeneous on large scales
can also be  phrased in terms of the fractal structure of the 
universe.
A fractal is a geometric shape that is not homogeneous, 
yet preserves the property that each part is a reduced-scale
version of the whole.
If the matter in the universe were actually 
distributed like a pure fractal on all scales then the 
Cosmological Principle 
would be invalid, and the standard model in trouble.
As shown in Figure 2 
current data already strongly constrain any non-uniformities in the 
galaxy distribution (as well as the overall mass distribution) 
on scales $> 300 \Mpc$.

If we count, for each galaxy,
the number of galaxies within a distance $R$ from it, and call the
average number obtained $N(<R)$, then the distribution is said to be a
fractal of correlation dimension $D_2$ 
if $N(<R)\propto R^{D_2}$. Of course $D_2$
may be 3, in which case the distribution is homogeneous rather than
fractal.  In the pure fractal model this power law holds for all
scales of $R$.

The fractal proponents (Pietronero et al. 1997)  have
estimated $D_2\approx 2$ for all scales up to $\sim 1000\Mpc$, whereas
other
groups 
have obtained scale-dependent values 
(for review see Wu et al. and references therein).

These measurements can be directly compared with the popular Cold Dark
Matter models of density fluctuations, which predict the increase
of $D_2$ with $R$ for the hybrid fractal model. If we now assume
homogeneity on large scales, 
then we have a direct mapping
between  correlation function $\xi(r)$ (or the Power-spectrum)  
and $D_2$. For 
$\xi(r) \propto r^{-\gamma}$
it follows that $D_2=3-\gamma$ if $\xi\gg 1$, while
if $\xi(r)=0$ then $D_2=3$. 
The predicted behaviour of  $D_2$ with $R$ 
from three different CDM models is shown 
Figure 3. 
Above $100\Mpc$ $D_2$  is 
indistinguishably close to 3. We also see  that it is
inappropriate to quote a single crossover scale to homogeneity, for
the transition is gradual.

\begin{figure}
\protect\centerline{
\psfig{figure=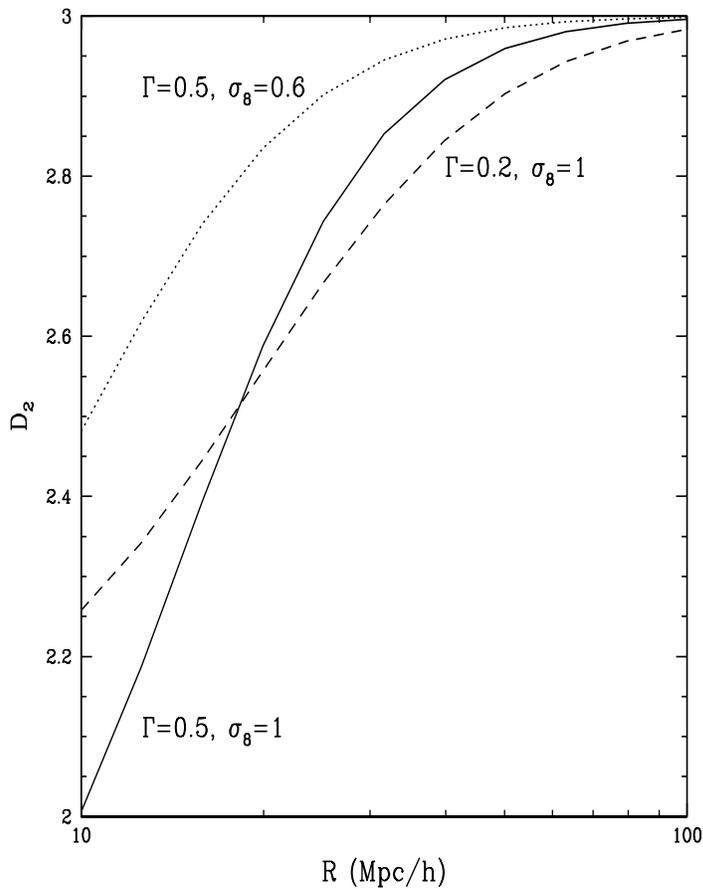,height=5truein,width=4truein}}
\caption[]{
The fractal correlation dimension $D_2$ versus
length scale $R$ assuming three Cold Dark Matter models of
power-spectra with shape and normalization parameters ($\Gamma =0.5$;
$\sigma_8=0.6$), ($\Gamma =0.5$; $\sigma_8=1.0$) and ($\Gamma =0.2$;
$\sigma_8=1.0$).  
They all exhibit the same qualitative behaviour of
increasing $D_2$ with $R$, becoming vanishingly close to 3 for $R >
100 h^{-1}$ Mpc
(from Wu, Lahav \& Rees 1998).}
\end{figure}

Direct estimates of $D_2$ are not possible for much larger scales, but
we can calculate values of $D_2$ at the scales probed by the XRB and
CMB by using CDM models normalised with the XRB and CMB as described
above.  The resulting values 
are consistent with $D_2=3$ to within
$10^{-4}$  on the very
large scales (Peebles 1993; Wu et al. 1998).
Isotropy does not imply homogeneity, but the near-isotropy of the CMB
can be combined with the Copernican principle that we are not in a
preferred position.  All observers would then  measure the same
near-isotropy, and an important result 
has been proven that 
the universe must then be very well approximated by 
the FRW metric (Maartens et al. 1996).

While we reject the pure fractal model in this review, the performance
of CDM-like models of fluctuations on large scales have yet to be
tested without assuming homogeneity {\it a priori}. On scales below,
say, $30\Mpc$, the fractal nature of clustering implies that one has
to exercise caution when using statistical methods which assume
homogeneity (e.g. in deriving cosmological parameters).  
As a final note, we emphasize that we only considered
one `alternative' here, which is the pure fractal model where $D_2$ is a
constant on all scales.

 \section {A `Best Fit Universe'}

Observations of anisotropies in the Cosmic Microwave Background (CMB)
provide one of the key constraints on cosmological models and a
significant quantity of experimental data already exists 
(e.g. Figure 2 and Gawiser \& Silk 1998)

On the other hand, 
galaxy redshift surveys, mapping large scale structure (LSS), provide
another cosmologically important set of observations.  The clustering
of galaxies in redshift-space is systematically different from that in
real-space (Kaiser 1987, Hamilton 1997 for review).
 The mapping between the two is a
function of the underlying mass distribution, in which the galaxies
are not only mass tracers, but also velocity test particles.
Estimates derived separately from each of the CMB and LSS  data sets have
problems with parameter degeneracy. 
Webster et al. (1998) combined results from a 
range of CMB experiments, with a likelihood analysis of the IRAS
1.2Jy survey, performed in spherical harmonics.  
This method expresses the effects of the
underlying mass distribution on both the CMB potential fluctuations
and the IRAS redshift distortion. This breaks the degeneracy inherent
in an isolated analysis of either data set, and places tight
constraints on several cosmological parameters. 

The family of CDM models analysed corresponds to a
spatially-flat universe with with an initially scale-invariant
spectrum and a cosmological constant $\lambda$. Free parameters in the joint
model are the mass density due to all matter ($\Omega$), Hubble's
parameter ($h = H_0 / 100$ km/sec), IRAS light-to-mass bias
($b_{iras}$) and the variance in the mass density field measured in an
$8 h^{-1}$ Mpc radius sphere ($\sigma_{8}$).  For fixed baryon density
$\Omega_b = 0.024/h^2$ the joint optimum lies at 
(Webster et al. 1998; Bridle et al., in preparation) 
$\Omega= 1 - \lambda
= 0.41\pm{0.13}$, $h = 0.52\pm{0.10}$, $\sigma_8 = 0.63\pm{0.15}$,
$b_{iras} = 1.28\pm{0.40}$ (marginalised 1-sigma error bars).
For these values of $\Omega, \lambda$ and $H_0$
the age of the universe is $\sim 16.6$ Gyr.

 \section {Discussion}

 We have shown some recent studies 
of  galaxy  surveys, and their cosmological implications. 
New measurements of galaxy clustering and background
radiations can provide improved constraints on the isotropy and
homogeneity of the Universe on large scales.  In
particular, the angular distribution of radio sources and the X-Ray
Background probe density fluctuations on scales intermediate between
those explored by galaxy surveys and CMB 
experiments.  On scales larger than $300 h^{-1} $ Mpc the distribution of
both mass and luminous sources satisfies well the `Cosmological
Principle' of isotropy and homogeneity.  Cosmological parameters 
such as $\Omega$ therefore have a well defined meaning.
 With the dramatic increase of data, we should soon be able to map
the fluctuations with scale and epoch, and to analyze jointly LSS 
(2dF, SDSS) and
CMB (MAP, Planck) data.

\bigskip

 { \bf Acknowledgments} 
 I thank my collaborators for their contribution to the work
 presented here. I also acknowledge  JSPS and 
 Tokyo and Kyoto Universities for the hospitality.


\section{References}

\vspace{1pc}


\re 
1. Baleisis, A., Lahav, O., Loan, A.J. and  Wall, J.V. 1998, 
{\it MNRAS}, {\bf 297}, 545.

\re
2. Baugh C.M. and  Efstathiou G., 1994, {\it MNRAS} , {\bf 267}, 323.

\re
3. Boldt, E. A., 1987,  {\it Phys. Reports}, {\bf 146}, 215.

\re
4. Cress C.M., Helfand D.J., Becker R.H., Gregg. M.D. and  White, R.L.,
1996,  {\it ApJ},  {\bf 473}, 7. 

\re
5. Dekel, A., 1994, {\it  ARAA}, {\bf 32}, 371.

\re
6. Fabian, A. C. and Barcons, X., 1992,  {\it ARAA}, {\bf 30}, 429.

\re
7. Folkes, S., Lahav, O. and   Maddox, S.J., 1996 {\it  MNRAS}, {\bf 283}, 651.

\re 
8. Folkes, S., Ronen, S., Price, I., Lahav, O., 
Colless, M.C., Maddox, S.J., Bland-Howthorn, J., Cannon, R., 
Cole, S., Collins, C., Couch, W., Driver, S.P.,
Dalton, G., Efstathiou, G., Ellis, R.S., Frenk, C.S., 
Glazebrook, K., Kaiser, N., Lewis, I., Lumsden, S., Peacock, J.A., 
Peterson, B., Sutherland, W. and Taylor, K., 1998, {\it in preparation}. 

\re
9. Gawiser, E. and Silk, J.,  1998, {\it Science}, {\bf 280}, 1405.

\re
10. Gunn, J.E. and  Weinberg, D.H., 1995,  
           in {\it Wide-Field Spectroscopy and the Distant Universe}, 
           eds. S.J. Maddox \& A. Aragon-Salamanca, World Scientific, 
\re
11. Hamilton , A. J. S., 1997, 
in {\it Ringberg Workshop on Large-Scale Structure}, 
Hamilton, D. (ed.), Kluwer Academic, Dordrecht,  astro-ph/9708102.

\re
12. Kaiser N., 1984,  {\it ApJ} , {\bf 284}, L9.

\re
13. Kochanek, C.S., 1996, {\it  ApJ}, {\bf 466}, 638.

\re
14. Lahav O., Piran T. and Treyer M.A.,  1997,  {\it MNRAS}, {\bf 284}, 499.

\re 
15. Maartens, R., Ellis, G. F. R. and Stoeger, W. R,
1996, {\it A\&A} , {\bf 309}, L7.

\re
16. Magliocchetti, M.,  Maddox, S.J., Lahav, O. and   Wall, J.V., 1998, 
{\it MNRAS}, in press, astro-ph/9802269.

\re 
17. Peebles, P. J. E., 1993, {\it  Principles of Physical Cosmology},
Princeton University Press, Princeton.

\re
18. Perlmutter, S., et al., 1998,  {\it Nature}, {\bf 391}, 51. 

\re
19. Pietronero, L., Montuori M., and Sylos-Labini, F., 1997, in 
{\it Critical Dialogues in Cosmology}, ed. N. Turok, astro-ph/9611197.

\re
20. Strauss M.A. and Willick J.A., 1995, {\it  Phys. Rep.},
 {\bf 261}, 271.

\re
21. Treyer, M., Scharf, C., Lahav, O., 
Jahoda, K.,  Boldt, E. and Piran, T., 1998,  
submitted to {\it ApJ}, astro-ph/9801293.

\re
22. Webster, M., Hobson, M.P., Lasenby, A.N., 
Lahav, O.,  Rocha, G. \& Bridle, S., 1998, 
submitted to {\it ApJL}, astro-ph/9802109.

\re
23. White, S.D.M., Navarro, J.F., Evrard, A.E. and  Frenk, C.S., 1993, 
{\it Nature}, {\bf 366}, 429.

\re
24. Wu, K.K.S.,  Lahav, O. \&   Rees, M.J., 1998,  submitted to {\it Nature}, 
astro-ph/9804062.

\vspace{1pc}

\end{document}